\documentclass[reqno,10pt,centertags]{amsart}
\usepackage{amsmath,amsthm,amscd,amssymb,latexsym,upref}

\newcommand{\R}{{\mathbb{R}}}

\newcommand{\calD}{{\mathcal D}}

\newcommand{\lb}{\label}

\newcommand{\g}{\gamma}
\newcommand{\Wo}{\stackrel{\scriptstyle\circ}{W_2^1}}

\newcommand{\sub}{\subset}
\newcommand{\Int}{\int\limits}
\newcommand{\Sup}{\sup\limits}

\newcommand{\Sum}{\sum\limits}

\renewcommand{\d}{\partial}
\renewcommand{\>}{\rangle}

\renewcommand{\a}{\alpha}
\renewcommand{\b}{\beta}

\allowdisplaybreaks
\numberwithin{equation}{section}

\newtheorem{theorem}{Theorem}[section]
\newtheorem{proposition}{Proposition}[section] 
\newtheorem{lemma}{Lemma}[section] 
\newtheorem{corollary}{Corollary}[section] 
\theoremstyle{definition}

\theoremstyle{remark}
\newtheorem{remark}[theorem]{Remark}

\begin{document}
\title{An inverse problem for an abstract evolution equation}
\author{S.~V.~Koshkin, A.~G.~Ramm}
\address{Department of Mathematics,
Kansas State University
Manhattan, Kansas,
66506-2602, USA}
\email{koshkin@math.ksu.edu; corresponding author: ramm@math.ksu.edu }

\begin{abstract} The inverse problem of finding the coefficient $\g$ 
in the equation $\dot{u}=A(t)u+\g(t)u+f(t)$ from the extra data of the form 
$\phi(t)=\<u(t),w\>$ is studied. The problem is reduced to a Volterra equation 
of the second kind. Applications are given to parabolic equations with second 
order differential operators.
\end{abstract}

\maketitle

\section{Introduction}\lb{s1}

Consider the Cauchy problem
\begin{equation}\lb{Cauch}
\dot{u}=A(t)u+\g(t)u+f(t),\ u(0)=u_0, 
\end{equation}
with the following extra data
\begin{equation}\lb{extra}
\phi(t)=\<u(t),w\>.
\end{equation}
Here $A(t)$ is a family of closed densely defined operators on a Banach space 
$X$, which generates an evolution family $U(t,s)$ (see \cite{Dal}), $u_0\in X$, 
$w\in X^*$ and $f$ is a given function on $[0,T]$ with values in $X$. The 
problem (\ref{Cauch}) is understood in the mild sense, i.e as solving the 
integral equation 
\begin{equation}\lb{ICauch}
u(t)=V(t,0)u_0+\Int_0^tV(t,s)f(s)ds,
\end{equation}
where $V(t,s)$ is the evolution family generated by $A(t)+\g(t)$. 
Here $\g(t)$ is an unknown scalar function of time, which should be recovered 
from 
the knowledge of the extra data $\phi(t)$. In applications $\g(t)$ may be a 
control function and $\phi(t)$ is a measured quantity. That is we have an 
inverse problem 
of finding control which brings the desired measurments.

Problems of this type when $A(t)$ is a second order differential operator with 
the Dirichlet boundary condition were studied in \cite{Kam}, \cite{PriS}. In 
\cite{Kam} a local existence result was obtained. In \cite{PriS} the authors 
assume that 
$w$ 
is a Dirac measure supported at some interior point of the domain and use
a complicated procedure to reduce the problem to the Volterra equation
similar to the one we derive in the general setting.
Our approach is similar to the one in \cite{LR}. In
\cite{R1} one can find a very general approach to many inverse problems,
the approach based on property C, that is, on completeness of the set of
products
of solutions to homogeneous differential equations. 

Let us describe the idea of the solution. One checks that
\begin{equation}\lb{V}
V(t,s)= e^{\Int_s^t\g(\tau)d\tau}U(t,s).
\end{equation}
Denote 
\begin{equation}\lb{xi}
\xi(t):= e^{-\Int_0^t\g(\tau)d\tau}
\end{equation}
and observe that 
$$
e^{\Int_s^t\g(\tau)d\tau}=\frac{\xi(s)}{\xi(t)}.
$$
Substituting (\ref{V}) and (\ref{xi}) into (\ref{ICauch}), one gets
\begin{equation}\lb{ut}
u(t)=\frac1{\xi(t)}U(t,0)u_0+\Int_0^t\frac{\xi(s)}{\xi(t)}U(t,s)f(s)ds.
\end{equation}
Now apply both sides of this equation to $w$ and multiply by $\xi(t)$ to get
\begin{equation}\lb{Voltr}
\phi(t)\xi(t)=\<U(t,0)u_0,w\>+\Int_0^t\<U(t,s)f(s),w\>\xi(s)ds.
\end{equation}
Note that in equation (\ref{Voltr}) all the functions are known except
$\xi(t)$. 
Assuming that $\phi(t)$ is separated from $0$ on $[0,T]$ and 
dividing both sides by this function, 
one gets a Volterra equation of the second kind for $\xi(t)$. Once $\xi(t)$ is 
found, $\g(t)$ can be found by the formula 
\begin{equation}\lb{gxi}
\g=-\frac{\dot{\xi}}{\xi}.
\end{equation}
Here one assumes that $\xi(t)$ corresponds to
some $\g$. Otherwise, (\ref{gxi}) implies (\ref{xi}) only under the 
additional conditions $\xi>0$, $\xi(0)=1$. This brings some additional 
restrictions on the coefficients of (\ref{Voltr}) to guarantee the existence of 
a solution to the inverse problem.

Section \ref{s2} contains precise formulations and proofs of our results and 
some of their applications.

\section{Solvability of the inverse problem}\lb{s2}

The uniqueness  result follows directly from (\ref{Voltr}) under very mild 
conditions. In what follows $\triangle_T:=\{(s,t)|0\leq s\leq t\leq T\}$.
By $c$ we denote below various positive constants.
\begin{proposition} In (\ref{Cauch})-(\ref{extra}) assume that  
\begin{enumerate}

\item $\|U(t,s)\|\leq M$ on $\triangle_T$. 

\item $f\in L^\infty([0,T],X)$ and $w\in X^*$.

\item $\phi(t)$ is measurable on $[0,T]$ and $\phi(t)\geq c>0$.

\end{enumerate}
Then there exists at most one solution $\g\in L^2([0,T])$ to the inverse 
problem 
(\ref{Cauch})-(\ref{extra}).
\end{proposition}
\begin{proof} If there are $\g_1$ and $\g_2$ such that (\ref{Cauch})-
(\ref{extra}) are 
satisfied then $\xi_1$ and $\xi_2$ defined by (\ref{xi}) will satisfy Volterra 
equation.  It follows from our assumptions that 
\begin{gather*}
\Sup_{t\in[0,T]}|\<U(t,0)u_0,w\>| \leq M\|u_0\|\|w\|<\infty,\\
\Sup_{(s,t)\in\triangle_T}|\<U(t,s)f(s),w\>| \leq M\|f\|\|w\|<\infty,
\end{gather*}
 so the kernels of the Volterra equation are bounded. They remain bounded after 
the division by $\phi(t)\geq c>0$ which turns (\ref{Voltr}) into a
Volterra equation 
of 
the second kind. Since this equation has at most 
one solution in $L^2([0,T])$ we 
conclude that $\xi_1=\xi_2$. Then $\g_1=\g_2$ from (\ref{gxi}).
\end{proof} 
One can prove the uniqueness of the solution of the Volterra integral
equation 
$h(x)=\int_0^x v(x,y)h(y)dy +f$ in $L^1([0,T])$ also in the case
when the kernel $v(x,y)$ is unbounded, for
example, if $\sup_{x\in [0,b]} \int_0^b|v(x,y)|dy<c.$
Existence results for the inverse problem are based on solving equation 
(\ref{Voltr}). But we need more from 
the solution than just being in $L^2$. 
The following lemma states conditions on the coefficients that guarantee these 
additional properties. 
\begin{lemma}\lb{lm} Consider the equation
\begin{equation}\lb{Vol}
\phi(t)\xi(t)=\a(t)+\Int_0^t\b(t,s)\xi(s)ds,
\end{equation}
where
\begin{enumerate}

\item $\a,\phi\in W_2^1([0,T])$, $\a(t)\phi(t)>0$ on $[0,T]$,

\item $\b\in L^\infty(\triangle_T)$, $\dot{\b}:=\frac{\partial\b}{\partial t}\in 
L^2(\triangle_T)$.

\newcounter{mycounter}
\setcounter{mycounter}{\value{enumi}}
\end{enumerate}
Then there exists a unique solution $\xi\in L^2([0,T])$ to (\ref{Vol}) and also 
$\xi(0)=\frac{\a(0)}{\phi(0)}$, $\xi\in W_2^1([0,T])$. Moreover, there exists 
$0<\tau\leq T$ such that $\xi(t)>0$ on $[0,\tau]$. If additionally
\begin{enumerate}
\setcounter{enumi}{\value{mycounter}}
\item $\beta\geq0$ on $\triangle_T$, $\phi(t)>0$ on $[0,T]$ 
\end{enumerate}
then $\xi(t)>0$ on $[0,T]$.
\end{lemma}
Proof of this lemma is standard (use iterations), and we omit it. 
\begin{proposition}\lb{loc} In (\ref{Cauch})-(\ref{extra}) assume that  
\begin{enumerate}

\item $\|U(t,s)\|\leq M$ on $\triangle_T$. 

\item $\phi\in W_2^1([0,T])$ and $\phi(0)=\<u_0,w\>\neq0$,

\item $\<U(t,0)u_0,w\>\in W_2^1([0,T])$, $\frac{\d}{\d t}\<U(t,s)f(s),w\>\in 
L^2(\triangle_T)$.
\end{enumerate}
Then there exist a $\tau$: $0<\tau\leq T$, and a unique pair $u\in 
C([0,\tau],X)$ and $\g\in L^2([0,\tau])$ which satisfies (\ref{Cauch})-
(\ref{extra}).
\end{proposition}
\begin{proof} Set $\a(t):=\<U(t,0)u_0,w\>$ and $\b(t,s):=\<U(t,s)f(s),w\>$. 
Since $\a(0)=\<U(0,0)u_0,w\>=\<u_0,w\>=\phi(0)\neq0$ and both functions $\a, 
\phi$ are continuous they have the same sign in some neighbourhood of $0$. So 
the conditions of Lemma \ref{lm} are satisfied in this case. Let $\xi$ be the 
solution of (\ref{Vol}) then $\xi\in W_2^1$, $\xi>0$ and $\xi(0)=\frac{\a(0)} 
{\phi(0)}=1$. Therefore, $\g$ defined by (\ref{gxi}) lies 
in $L^2([0,\tau])$ and 
satisfies (\ref{xi}). Define $u(t)$ by (\ref{ut}) then (\ref{Voltr}) says that 
$\phi(t)=\<u(t),w\>$ and (\ref{extra}) is satisfied. Finally, (\ref{xi}) 
together with (\ref{ut}) are equivalent to (\ref{V}) which means that 
(\ref{Cauch}) is satisfied also. Thus,
 $\g$ and $u$ solve the inverse problem on $[0,\tau]$.
\end{proof} 
\begin{proposition}\lb{glob} Let the conditions of Proposition \ref{loc} 
hold and in 
addition $\phi(t)>0$, $\<U(t,0)u_0,w\> >0$ on $[0,T]$ and 
$\<U(t,s)f(s),w\>\geq0$ on $\triangle_T$. Then there exists a unique pair 
$u\in C([0,T],X)$ and $\g\in L^2([0,T])$ which satisfies
\ref{Cauch}-\ref{extra}.
\end{proposition}
\begin{proof} The argument is essentially the same as above with one 
difference:  by Lemma \ref{lm} $\xi$ is now positive on the whole 
interval $[0,T]$ and not only in a small neighborhood of $0$ so that $\tau=T$.
\end{proof} 
\begin{remark}\lb{dual} Let $\calD_A$ be the domain of a closed densely defined 
linear operator $A$ and for a family $A(t)$ of such operators 
$\calD:=\cap_{t>0}\calD_{A(t)}$ with the corresponding norm 
$\|x\|_{\calD}:=\|x\|+\Sup_{t>0}\|x\|_{\calD_{A(t)}}$. Likewise, 
$\calD^*:=\cap_{t>0}\calD_{A^*(t)}$. If $w\in\calD^*,$ then some 
smoothness conditions are satisfied. In particular,
$$
\frac{d}{dt}\<U(t,0)u_0,w\>=\<u_0,\frac{d}{dt}U^*(t,0)w\>=
\<u_0,U^*(t,0) A^*(t)w\>\in L^\infty([0,T])  
$$
$$
\frac{d}{dt}\<U(t,s)f(s),w\>=\<f(s),\frac{d}{dt}U^*(t,s)w\>=
\<f(s),U^*(t,s)A^*(t)w\>\in L^\infty([0,T]).
$$
\end{remark}

Consider the problem of finding a coefficient 
of a parabolic equation. Let $\Omega\sub\R^n$ be a bounded domain 
with Lipschitz  boundary, define $X:=L^2(\Omega)$. Then the
family of uniformly elliptic operators $A(t)$ 
may be defined by the differential expression 
\begin{equation}\lb{parop}
\frac{\d}{\d x_i}(a_{ij}(x,t)\frac{\d}{\d x_j})+b_i(x,t)\frac{\d}{\d x_i}+a(x,t)
\end{equation}
(summation over the repeated indices is understood, $1\leq i,j \leq
3$) and
the Dirichlet boundary 
conditions. Under certain restrictions on the coefficients it follows from the 
general theory of parabolic equations (see \cite{Lad}) that the Cauchy problem
\begin{equation} 
\dot{u}=A(t)u,\ u(0)=u_0,
\end{equation}
is uniquely solvable in $C([0,T],L^2(\Omega))$. Therefore, it defines an 
evolution family $U(t,s)$ by setting $U(t,s)u_0$ equal to the solution of
$$
\dot{u}=A(t)u,\ u(s)=u_0,\ s\leq t
$$ 
at the moment $t$. Using Proposition \ref{loc} we can easily establish the
result 
proved for the one dimensional case, $\Omega=[0,1]$ in \cite{Kam}.
 The closure in the norm of the Sobolev space $W^1_2(\Omega)$ of the set
of smooth and compactly supported in $\Omega$
functions is denoted by $\Wo(\Omega)$. By $W_{2,0}^2(\Omega)$ we denote
the intersection of $W^2_2(\Omega)$ and  $\Wo(\Omega)$.

\begin{corollary}\lb{parloc} Consider the problem (\ref{Cauch})-(\ref{extra}), 
where $A(t)$ is defined by (\ref{parop}). Assume that 

\begin{enumerate}

\item\lb{par1} $\exists \mu,\nu>0: \nu\Sum_i\xi^2_i\leq 
a_{ij}(x,t)\xi_i\xi_j\leq\mu\Sum_i\xi^2_i$,

\item\lb{par2}  $a,b_i\in L^\infty([0,T],L^\infty(\Omega))$, $\frac{\d 
a_{ij}}{\d t}\in L^\infty([0,T],L^\infty(\Omega))$,

\item\lb{par3} $u_0\in \Wo(\Omega)$, $f\in L^\infty([0,T], 
\Wo(\Omega))$. 

\item\lb{par4} $w\in L_2(\Omega)$, $\phi\in W_2^1([0,T])$ and 
$\phi(0)=\<u_0,w\>\neq0$. 

\end{enumerate}

Then there exist a $\tau$: $0<\tau\leq T$, and a unique pair $u\in 
W^1_2([0,\tau],W^2_{2,0}(\Omega))$ and 
$\g\in L^2([0,\tau])$ which satisfies (\ref{Cauch})-(\ref{extra}).
\end{corollary} 
\begin{proof} Recall that $U(t,0)u_0$ is a solution to the parabolic problem 
with the initial data $u_0$. Under the conditions \ref{par1}.-\ref{par4}.
of Corollary \ref{parloc} it 
is known that $U(t,0)u_0\in W_2^1([0,T], W^1_2(\Omega))$
(\cite{Lad}, p.178-180). Then for 
$w\in L^2(\Omega)$ the function $\<U(t,0)u_0,w\>$ is in $W_2^1([0,T])$. 
Similarly, $\frac{\d}{\d t}U(t,s)f(s)\in
L^\infty([s,T],W^2_{2,0}(\Omega))$ for
a
fixed $s$ and, therefore, 
$\frac{\d}{\d t}\<U(t,s)f(s),w\>\in L^2(\triangle_T)$. 
Thus, all of the conditions of Proposition \ref{loc} are fulfiled.
\end{proof}
An alternative reference to \cite{Lad} may be \cite{Paz},
where Theorem 6.1 on p.150 shows that our operator $A(t)$
generates an evolution family in $L^2(\Omega)$.

For the global solvability it is important that $U(t,s)$ 
be a positive evolution 
family in the sense that $u_0\geq0$ implies $U(t,s)u_0\geq0$ (pointwise) due to 
the maximum principle for the parabolic equations \cite{Lad},\cite{Max}. The 
following result (Corollary 2.2) is similar to the one in \cite{PriS}. The
authors of \cite{PriS}, 
however, used different spaces and the data which correspond
in our general scheme to the choice of $w$ as 
a delta-function supported at an interior point of $\Omega$, see remark
2.2 below.
\begin{corollary}\lb{parglob} In addition to the
assumptions \ref{par1}.-\ref{par4}. of 
Corollary \ref{parloc} assume that 

$\phi,w>0$ on $[0,T]$; $u_0,f\geq0$, $u_0\neq0$.

\noindent Then there exists a unique pair $u\in C([0,T],L^2(\Omega))$
and  
$\g\in L^2([0,T])$ which satisfies (\ref{Cauch})-(\ref{extra}).
\end{corollary} 
\begin{proof} By Proposition \ref{glob} it suffices to prove that
$$
\<U(t,0)u_0,w\> >0, \<U(t,s)f(s),w\>\geq0.
$$ 
The latter follows from the usual maximum 
principle for parabolic equations (\cite{Lad}, p.188). For the same reason 
$\<U(t,0)u_0,w\>\geq0$. Assume that $\<U(t,0)u_0,w\>=0$ for some $t>0$. Since 
$w>0$ one obtains $U(t,0)u_0=0$ for some $t>0$. Then, by the strong
maximum 
principle (\cite{Max}, p.174-175), $u_0=0$ in condradiction with our assumption. 
Thus, 
$\<U(t,0)u_0,w\> >0$ on $[0,T]$ as we claimed.
\end{proof}
\begin{remark} Formally the case of measurements at one point, i.e. 
$w=\delta_{x_0},\ x_0\in\Omega$, is not covered by this theorem since 
$\delta_{x_0}\notin L_2(\Omega)$. However, it may be easily included in the 
general scheme by choosing the data into such spaces that 
the solution to the
parabolic problem 
belongs to $W_2^1([0,T],X)$ with $\delta_{x_0}\in X^*$. A possible 
choice is the H\H{o}lder spaces used in \cite{PriS}.
\end{remark}

The above constraints on $A(t), u_0,$ and $ f$ can be weakened 
if additional constraints are imposed 
on $w$ as explained in remark \ref{dual}.

\begin{corollary} In the problem (\ref{Cauch})-(\ref{extra}), where $A(t)$ is 
defined by (\ref{parop}), assume that 

\begin{enumerate}

\item $\exists \mu,\nu>0: \nu\Sum_i\xi^2_i\leq 
a_{ij}(x,t)\xi_i\xi_j\leq\mu\Sum_i\xi^2_i$,

\item $a,b_i\in L^\infty([0,T],L^\infty(\Omega))$, 

\item $f\in L^\infty([0,T],L^2(\Omega))$, $u_0\in L^2(\Omega)$, 

\item $w\in W_{2,0}^2(\Omega)$, 
$\phi\in W_2^1([0,T])$ and 
$\phi(0)=\<u_0,w\>\neq0$, 

\item $\phi,w>0$ on $[0,T]$; $u_0,f\geq0$, $u_0\neq0$.
\end{enumerate}

Then there exists a unique pair $u\in L^\infty([0,T])$, 
$\g\in L^2([0,T])$ which satisfies (\ref{Cauch})-(\ref{extra}).
\end{corollary} 
\begin{proof} It suffices to note that
$W_{2,0}^2(\Omega)\sub\calD^*$ and use 
remark \ref{dual} to verify the necessary smoothness. The rest of
argument is 
the same as in corollaries \ref{parloc},\ref{parglob}.
\end{proof}


\end{document}